\documentclass[amsmath,prb,twocolumn,showpacs,superscriptaddress]{revtex4}

\usepackage{amsfonts}
\usepackage{graphicx}
\usepackage{pstricks}

\begin{document}

\title{Inelastic scattering and heating in a molecular spin pump}
\date{\today}
\author{Jonas Fransson}
\affiliation{Department of Physics and Materials Science, Uppsala University,
Box 530, SE-751 21 Uppsala, Sweden}
\author{Michael Galperin}
\affiliation{Department of Chemistry \& Biochemistry, University of California 
at San Diego, La Jolla CA 92093, USA}

\begin{abstract}
We consider a model for a spin field-effect molecular transistor, where a directed pure spin current is controlled by an external electric field. Inelastic scattering effects of such molecular device are discussed within a framework of full counting statistics for a multi-level molecular system. We propose that the heating of the molecular junction can be controlled by external electric and magnetic fields. Characteristic features of the model are demonstrated by numerical calculations.
\end{abstract}

\pacs{85.65.+h 85.75.Hh 71.38.-k 73.20.Hb}

\maketitle
%%%%%%%%%%%%%%%%%%%%%%%%%%%%%%%%%%%%%%%%%%%%%%%%%%%%%%%%%%%%%%%%%%%%%%%%%%%%%%%
\section{Introduction}\label{intro}
Fast development of experimental techniques allowing for miniaturization of 
electronic devices led to renewed interest in theoretical research in 
the area of quantum transport. In particular in molecular electronics
community focus of the research was shifting from elastic (Landauer)
to inelastic charge transport through molecular junction, to its noise 
characteristics and heating. Spin flux as an alternative to charge 
current in electronic devices, 
where magnetic field provides an additional (to bias) mean of control, 
is studied by spintronics.\cite{DasSarma_review,Kouwenhoven_review}
Quantum ratchets\cite{Flatte,Burkard} and electric 
potential\cite{Halperin,Gossard,Debray} 
were proposed as additional controls of the spin flux.
Schemes for optical control of a spin trapped in quantum dots were
reported in the literature.\cite{Awschalom,McEuen,Press,Paaske,Galand,Dubin}
Recently a combination of spintronics with molecular electronics
started to reveal itself as molecular 
spintronics.\cite{Ralph,Rocha,Sanvito,Wernsdorfer}
Theoretical schemes for spin pumps were considered in 
Refs.~\onlinecite{Guo_spinpump,PengChen,FranssonZhu_rf},
and inelastic effects of spin transport through the junctions
were reported for spin valves\cite{Sanvito_sv} and 
for tunneling through a junction with embedded spin.\cite{Fransson}
Shot noise of spin current was considered in Ref.~\onlinecite{Guo_noise}.

Here we present a model for a molecular junction, consisting
of a molecule between two normal metallic contacts, where pure
spin current is controlled by an external electric field.
The spin field-effect transistor is a generalization of a
spin pump introduced in Ref.~\onlinecite{Guo_spinpump}.
Directed spin current seems to be a more convenient element for
molecular spintronics device. As in the case of the spin pump, 
only pure spin current participates in the transport. We consider 
inelastic effects of the spin current revealed in the transport properties 
of the device. This consideration takes place as a part of a  general approach
of full counting statistics for multi-level molecular systems.
In particular, we discuss current and zero frequency noise of the junction.

Flexibility of molecules, and as a result well-pronounced inelastic features
in transport properties of molecular junctions, makes heating of a device
an important issue in molecular electronics. Spin field-effect transistors (FET)
inherit this problem of the usual FETs, since spin current
is also caused by electrons crossing molecule-contact interface 
which results in
heating molecular device. Within the model we discuss spin current
through molecular junction, and propose external electric and magnetic 
fields as possible controls capable of tuning molecular structure 
to diminish heating of the device.

Section \ref{model} presents a model and outlines the method.
Section \ref{fcs} is devoted to full counting statistics of
multi-level molecular systems. We describe general approach and 
introduce junction characteristics -- spin and charge current and noise.
Section \ref{heat} is devoted to heating in
spin field-effect transistors. In section \ref{numres} we present
results of numerical simulations. Section \ref{conclude} concludes.

%%%%%%%%%%%%%%%%%%%%%%%%%%%%%%%%%%%%%%%%%%%%%%%%%%%%%%%%%%%%%%%%%%%%%%%%%%%%%%%
\section{\label{model}Model}
We consider extension of the spin pump proposed in 
Ref.~\onlinecite{Guo_spinpump} to the junction situation. 
The spin pump of Ref.~\onlinecite{Guo_spinpump} is a model of a two-level system coupled to
one electrode. Application of magnetic field generates spin current
in the contact due to spin-flip process taking place at the molecule.
We propose a generalization where the spin flux is generated across the
junction (spin current between two contacts). Note that the spin flux
in our model is optically controlled, i.e. external electric rather
than magnetic field controls spin flux.
The junction is composed of two molecules
(represented by single levels $1$ and $2$ coupled to molecular vibrations
$\omega_1$ and $\omega_2$) attached to left $L$
and right $R$ normal metal electrodes, respectively. 
Each molecule is subject to a {\em dc} and {\em ac} pair of magnetic fields 
$B^{(dc)}_i$ and $B^{(ac)}_i$ ($i=1,2$). 
Note that level dependent g-factors in nanowire quantum dots were
recently reported in the literature,\cite{Nilsson} something that, here, is represented by the site dependent field.
The molecules are weakly coupled by external source of light of particular 
frequency $\omega_E$. The light can be used as a handle to switch on and off
spin transport through the junction. The Hamiltonian of the system is
\begin{align}
 \label{H}
 \hat H =& \hat H_0 + \hat V,
\end{align}
where
\begin{align}
 \label{H0}
 \hat H_0 =&
	\sum_{k\in L,R;\sigma} \varepsilon_{k} \hat n_{k\sigma}
 	+\sum_{i=1,2}\biggr(\sum_\sigma[\varepsilon_{i\sigma}+M_i(\hat a_i+\hat a_i^\dagger)]\hat n_{i\sigma}
 \nonumber \\
 &
		+\omega_i \hat a_i^\dagger\hat a_i
 	-g\mu_B B^{(ac)}_i 
		\Bigl(\hat d_{i\uparrow}^\dagger\hat d_{i\downarrow}e^{ i\omega^{(B)}_i t}
		+H.c.
		\Bigr)\biggr)
 \nonumber \\
 &
 	+\sum_{k\in L;\sigma}\Bigr(
		V_{1k}\hat d_{1\sigma}^\dagger\hat c_{k\sigma}+H.c.\Bigr)
\nonumber\\
&
	+\sum_{k\in R;\sigma}\Bigl(
		V_{2k}\hat d_{2\sigma}^\dagger\hat c_{k\sigma} 
		+H.c.\Bigl),
\end{align}
\begin{align}
 \label{V}
 \hat V =& (V_E e^{-i\omega_E t}+V_E^{*} e^{i\omega_E t})
           \sum_\sigma (\hat d_{1\sigma}^\dagger\hat d_{2\sigma}
	+H.c.).
\end{align}
where $d_{i\sigma}^\dagger$ ($\hat d_{i\sigma}$) and $\hat c_{k\sigma}^\dagger$
($\hat c_{k\sigma}$) are creation (annihilation) operators for
corresponding state in the molecule(s) and in the contact(s),
$\hat n_{i\sigma}=\hat d_{i\sigma}^\dagger\hat d_{i\sigma}$
and $\hat n_{k\sigma}=\hat c_{k\sigma}^\dagger\hat c_{k\sigma}$,
$\hat a_i^\dagger$ ($\hat a_i$) are creation (annihilation)
operator for vibration quanta of molecule $i$,
$\sigma=\pm 1$ is direction of spin projection, and where
\begin{equation}
 \varepsilon_{i\sigma} = \varepsilon_i - \frac{\sigma}{2}g\mu_B B^{(dc)}_i.
\end{equation}

\begin{figure}[tbp]
\centering\includegraphics[width=\linewidth]{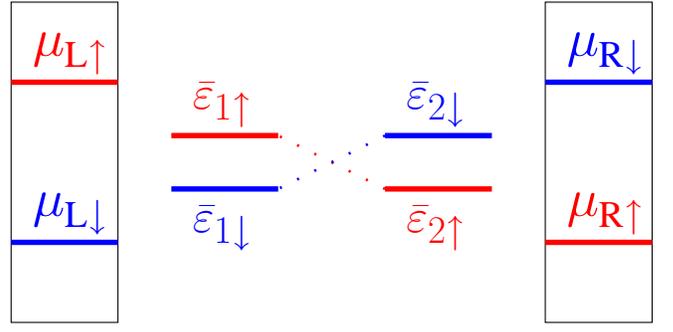}
\caption{\label{scheme}
(Color online)
Sketch of an optically controlled spin field-effect transistor.
}
\end{figure}

We start by transforming the Hamiltonian (\ref{H}) into the rotating frames of the  magnetic field \cite{FranssonZhu_rf,ZhagXuXie_rf}
\begin{align}
 \label{rf}
 \hat H \to& \hat{\bar H} = 
 i\left(\frac{\partial}{\partial t}e^{\hat S_B(t)}\right)
 e^{-\hat S_B(t)} + e^{\hat S_B(t)}\hat H e^{-\hat S_B(t)},
 \\
 S_B(t) =& -i\frac{\omega^{(B)}_1}{2} t \sum_\sigma\sigma
          \left(\hat n_{1\sigma}+\sum_{k\in L}\hat n_{k\sigma}\right)
 \nonumber \\
         & -i\frac{\omega^{(B)}_2}{2} t \sum_\sigma\sigma
          \left(\hat n_{2\sigma}+\sum_{k\in R}\hat n_{k\sigma}\right).
\end{align}
This transformation eliminates the time-dependence from the ac magnetic field terms,
shifts the positions of molecular levels, and induces spin biases in the contacts, according to
\begin{align}
 \label{eibar}
 \bar\varepsilon_{i\sigma} =& \varepsilon_{i\sigma} + \sigma{\omega^{(B)}_i}/{2},
 \\
 \label{ekbar}
 \bar\varepsilon_{k\sigma} =& \left\{
 \begin{array}{ll}
    \varepsilon_{k}+\sigma\omega^{(B)}_1/2 & k\in L, \\
    \varepsilon_{k}+\sigma\omega^{(B)}_2/2 & k\in R.
 \end{array}
 \right.
\end{align}
The perturbation (\ref{V}) takes the form
\begin{align}
 \label{Vbar}
 \hat{\bar V} =& (V_E e^{-i\omega_E t}+V_E^{*} e^{i\omega_E t})
 \nonumber \\ \times&
  \sum_\sigma \left(\hat d_{1\sigma}^\dagger\hat d_{2\sigma}
               e^{-i\sigma(\omega^{(B)}_1-\omega^{(B)}_2)t/2}
               +H.c.\right)
\end{align}
In what follows we will put $\omega^{(B)}_1=-\omega^{(B)}_2\equiv\omega_B$
to create a spin bias across the junction 
(see Fig.~\ref{scheme} for a sketch). 
Similarly, we put $g\mu_B B_1^{(dc)}=-g\mu_B B_2^{(dc)}\equiv\omega_0$.
The interaction with the external optical field is taken into account within
perturbation theory. Expansion of the evolution operator on the Keldysh 
contour up to second order in (\ref{Vbar}) in the rotating-wave 
approximation (RWA) leads to electronic self-energy due to interaction
with external optical field in the form (see Appendix~\ref{appA} for
details)
\begin{align}
 \label{SE}
 &\Sigma_{\sigma,\sigma'}^{(E)}(\tau,\tau') = \delta_{\sigma,\sigma'}
 2 |V_E|^2 \cos\omega_E(t-t') 
 \\ &\times
 \left[
 \begin{array}{cc}
 G_{2\sigma,2\sigma}(\tau,\tau') e^{-i\sigma\omega_B(t-t')} & 0 \\
 0 & G_{1\sigma,1\sigma}(\tau,\tau') e^{i\sigma\omega_B(t-t')}
 \end{array} 
 \right].
 \nonumber
\end{align}
As a result of the transformation to the rotating frames and the RWA we obtain a
time-independent (steady-state) description. 

The coupling between electronic and vibrational degrees of freedom
is treated below within two different approaches.
The transport through the junction is considered in a basis
obtained by a small polaron transformation\cite{Mahan}
\begin{align}
 \label{sp}
 \hat H \to& \hat{\bar H} = e^{\hat S_v}\hat H e^{-\hat S_v}
 \\
 \hat S_v =& \sum_{i=1,2;\sigma}\frac{M_i}{\omega_i}(\hat a_i^\dagger-\hat a_i)
 \hat n_{i\sigma}.
\end{align}
This transformation
decouples the electronic and vibrational degrees of freedom on the molecule
and dresses the molecular fermion operators
\begin{equation}
 \label{diX}
 \hat d_{i\sigma} \to \hat d_{i\sigma}\hat X_i
 \qquad \hat X_i = 
 \exp\left[-\frac{M_i}{\omega_i}(\hat a_i^\dagger-\hat a_i)\right].
\end{equation}
We take the vibrational shift operators $\hat X$ into account within 
the usual Born-Oppenheimer-like approximation, which allows us to introduce the
Franck-Condon factors\cite{CK_inel} multiplying the electronic
GF
\begin{align}
 \label{GF_FC}
 &G_{i\sigma,j\sigma'}(\tau,\tau') \equiv 
 -i\langle T_c \hat d_{i\sigma}(\tau)\hat X_i(\tau)\,
        \hat d_{j\sigma'}(\tau')\hat X_j(\tau')\rangle
 \nonumber \\
 &\approx -i\langle T_c \hat d_{i\sigma}(\tau)\,\hat d_{j\sigma'}^\dagger(\tau')\rangle
           \langle T_c \hat X_i(\tau)\,\hat X_j^\dagger(\tau')\rangle
\end{align}

In the description of the junction heating, we utilize the Born 
approximation\cite{peaksdips} instead, which allows us to keep 
the consideration simple. As usual we implement a non-crossing
approximation, i.e. diagrams for electron transfer and 
interaction with vibrations do not cross (the processes do not
happen simultaneously).\cite{Mahan}
In this case Green function is obtained from the Dyson equation
\begin{align}
 \label{GF_BA}
 G_{i\sigma,j\sigma'}(\tau,\tau') &= G^{(0)}_{i\sigma,j\sigma'}(\tau,\tau')
 \\
 &+ \sum_{m_1,m_2}\sum_{\sigma_1,\sigma_2}\int_c d\tau_1\int_c d\tau_2\,
   G^{(0)}_{i\sigma,m_1\sigma_1}(\tau,\tau_1)\,
 \nonumber\\ &\times
   \Sigma^{(ph)}_{m_1\sigma_1,m_2\sigma_2}(\tau_1,\tau_2)\,
   G_{m_2\sigma_2,j\sigma'}(\tau_2,\tau')
 \nonumber
\end{align}
where $G^{(0)}$ is Green function in the absence of electron-phonon
interaction ($M_{1,2}=0$), and $\Sigma^{(ph)}$ is self-energy due to
electron-phonon interaction
\begin{equation}
 \label{SEph}
 \Sigma^{(ph)}_{i\sigma,j\sigma'}(\tau,\tau')
 = i \delta_{i,j} |M_i|^2 D_i(\tau,\tau')\, G_{i\sigma,i\sigma'}(\tau,\tau')
\end{equation}
Here $D_i(\tau,\tau')\equiv -i\langle T_c \hat a_i(\tau)\hat a_i^\dagger(\tau')\rangle$
is phonon Green function. We treat it within a quasi-particle
approximation.\cite{KadanoffBaym}

%%%%%%%%%%%%%%%%%%%%%%%%%%%%%%%%%%%%%%%%%%%%%%%%%%%%%%%%%%%%%%%%%%%%%%%%%%%%%%%
\section{\label{fcs}Full counting statistics}
The theoretical concept of full counting statistics was originally proposed 
by Levitov and Lesovik.\cite{LevitovLesovik_JETP,LevitovLesovik_JMP} 
The approach was applied to the non-equilibrium Anderson impurity model
in Ref. \onlinecite{GogolinKomnik}. Measurements of shot noise in 
molecular junction\cite{Ruitenbeek} prove possibility of experimental 
observation of moments beyond average current. This together with the
flexibility of molecules, i.e. importance of inelastic effects in
transport through molecular junctions, recently caused several
theoretical investigations devoted to study of resonant level
coupled to single vibration model.\cite{Komnik,Yeyati,Belzig}

Here we discuss a generalization of the result derived 
by Gogolin and Komnik\cite{GogolinKomnik} to a multilevel situation,
and apply the expression to calculate the first (current) and second (noise)
cumulants of the distribution. Following the derivation of 
Ref.~\onlinecite{GogolinKomnik} for the case of multilevel molecule
we obtain an expression for the derivative of the adiabatic potential in 
the form\cite{footnote_1}
\begin{widetext}
\begin{align}
 \label{dU}
 &\frac{\partial}{\partial\lambda_K^{-}}\mathcal{U}(\lambda_K^{-},\lambda_K^{+})
 =-\frac{1}{2}\int_{-\infty}^{+\infty}\frac{dE}{2\pi}
  \mbox{Tr}\left\{
   \mathbf{\Sigma}_{K}^{<}(E)e^{i\lambda_K} \mathbf{G}_{\lambda}^{>}
  -\mathbf{G}_{\lambda}^{>}\mathbf{\Sigma}_{K}^{<}(E)e^{-i\lambda_K}
  \right\}
 \equiv -\frac{1}{2}\int_{-\infty}^{+\infty}\frac{dE}{2\pi}
 \nonumber \\
 &\mbox{Tr}\left\{
 \mathbf{\Sigma}_{K}^{<}(E)e^{i\lambda_K}
 \left[\left(E-\mathbf{H}_0-\mathbf{\Sigma}^{--}(E)\right)
       \left[\mathbf{\Sigma}_{\lambda}^{>}(E)\right]^{-1}
       \left(E-\mathbf{H}_0+\mathbf{\Sigma}^{++}(E)\right)
       +\mathbf{\Sigma}_{\lambda}^{<}(E)
 \right]^{-1}
 \right. \\
 &\left.\ \  -
 \left[\left(E-\mathbf{H}_0+\mathbf{\Sigma}^{++}(E)\right)
       \left[\mathbf{\Sigma}_{\lambda}^{<}(E)\right]^{-1}
       \left(E-\mathbf{H}_0-\mathbf{\Sigma}^{--}(E)\right)
       +\mathbf{\Sigma}_{\lambda}^{>}(E)
 \right]^{-1}
 \mathbf{\Sigma}_K^{>}e^{-i\lambda_K}
 \right\}.
 \nonumber
\end{align}
\end{widetext}
Here, $\lambda_K^{-}$ ($\lambda_K^{+}$) is a counting field 
for the interface between the molecule and contact $K$ on the 
forward (backward) branch of the Keldysh contour, 
$\lambda_K\equiv(\lambda_K^{-}-\lambda_K^{+})/2$. 
`$--$', `$++$', `$<$', and `$>$' are causal, anti-causal, lesser,
and greater projections.
The trace runs over molecular degrees of freedom.
The electronic self-energy due to the coupling to contact $K$ is denoted by
$\mathbf{\Sigma}_K$, and $\mathbf{\Sigma}_{\lambda}$ is the 
total electronic self-energy dressed with
the counting field $\lambda$. In particular, within the non-crossing
approximation\cite{Mahan}
\begin{equation}
 \mathbf{\Sigma}^{>,<}_{\lambda}(E) = \sum_K \mathbf{\Sigma}^{>,<}_K(E)
 e^{\mp i\lambda_K} + \mathbf{\Sigma}^{>,<}_{int,\lambda}(E),
\end{equation}
where $\mathbf{\Sigma}^{>,<}_{int,\lambda}$ is the electronic self-energy
due to interactions dressed with the counting field 
$\lambda$.

It seems difficult to obtain an expression for the logarithm
of the generating function (integral of (\ref{dU}) over $\lambda_K^{-}$)
in the multi-level case. However, Eq. (\ref{dU}) itself can be used to calculate
cumulants. The time-averaged charge cumulant of order $n$ due to charge transport 
through interfaces $\{K_j\}$ ($j=\{1,2,\ldots,n\}$) is
\begin{align}
 &\frac{\langle\delta^n q\rangle_{K_n,\ldots,K_1}}{T} = 
 \\
 &\quad -i \frac{\partial}{\partial(i\lambda_{K_n})}\ldots
           \frac{\partial}{\partial(i\lambda_{K_1})}
 \left.\mathcal{U}(\{\lambda\},-\{\lambda\})\right|_{\{\lambda\}=0}.
 \nonumber
\end{align}
Here and below $e=\hbar=1$.

First cumulant yields the well-know expression for 
steady-state current\cite{MeirWingreen,HaugJauho}
\begin{align}
 \label{IK}
 I_K =& -i\frac{\partial}{\partial(i\lambda_K)}
 \left.\mathcal{U}(\{\lambda\},-\{\lambda\})\right|_{\{\lambda\}=0}
 \\
 =& \int_{-\infty}^{+\infty}\frac{dE}{2\pi}\,
   \mbox{Tr}\left[\mathbf{\Sigma}_K^{<}(E)\,\mathbf{G}^{>}(E)
                 -\mathbf{\Sigma}_K^{>}(E)\,\mathbf{G}^{<}(E)\right].
 \nonumber
\end{align}
Since no spin-flip events are allowed on the metal-molecule interface, the electronic self-energy due to coupling to the contacts is diagonal in spin space, that is,
\begin{equation}
 \label{SK}
 \Sigma_\sigma^{(K)}(\tau,\tau') = 
 \sum_{k\in K} |V_{ik}|^2 g_{k\sigma}(\tau,\tau'),
\end{equation}
where $i=1$ for $K=L$ and $i=2$ for $K=R$, whereas $g_{k\sigma}$
is the Green function (GF) of a free electron. As a result, the charge current of electrons with spin $\sigma$
at interface $K$ is given by
\begin{align}
 \label{IKs}
 I_{K\sigma} =& \frac{e}{\hbar}\int_{-\infty}^{+\infty}\frac{dE}{2\pi}\,
 \\
 &\left[\Sigma_\sigma^{(K)<}(E)\,G_{i\sigma,i\sigma}^{>}(E)
       -\Sigma_\sigma^{(K)>}(E)\,G_{i\sigma,i\sigma}^{<}(E)\right].
 \nonumber
\end{align}
The spin and charge currents at interface $K$ are
\begin{align}
 I^{(s)}_K =& I_{K\uparrow} - I_{K\downarrow}, \\
 I^{(c)}_K =& I_{K\uparrow} + I_{K\downarrow},
\end{align}
respectively. 

The second cumulant yields an expression for the zero-frequency noise.
In the non-interacting model, it reads\cite{footnote_2} 
\begin{widetext}
\begin{align}
 \label{SK1K2}
 &S_{K_2K_1}(\omega=0) = -i\frac{\partial}{\partial(i\lambda_{K_2})}
 \frac{\partial}{\partial(i\lambda_{K_1})}
 \left.\mathcal{U}(\{\lambda\},-\{\lambda\})\right|_{\{\lambda\}=0}
 \nonumber \\
 &= \int_{-\infty}^{+\infty}\frac{dE}{2\pi}\,\mbox{Tr}\left\{
 \delta_{K_1,K_2}\left[\mathbf{\Sigma}_{K_1}^{<}(E)\,\mathbf{G}^{>}(E)
                      +\mathbf{G}^{<}(E)\,\mathbf{\Sigma}_{K_2}^{>}(E)\right]
 -\mathbf{i}_{K_1}(E)\,\mathbf{i}_{K_2}(E)
 \right. \nonumber \\ &\qquad\qquad\qquad
 +\mathbf{\Sigma}_{K_1}^{<}(E)\,\mathbf{G}^{++}(E)\,
  \mathbf{\Sigma}_{K_2}^{>}(E)\,\mathbf{G}^{--}(E)
 +\mathbf{\Sigma}_{K_1}^{>}(E)\,\mathbf{G}^{--}(E)\,
  \mathbf{\Sigma}_{K_2}^{<}(E)\,\mathbf{G}^{++}(E)
 \\ &\left.\qquad\qquad\qquad
 -\mathbf{\Sigma}_{K_1}^{<}(E)\,\mathbf{G}^{>}(E)\,
  \mathbf{\Sigma}_{K_2}^{>}(E)\,\mathbf{G}^{<}(E)
 -\mathbf{\Sigma}_{K_1}^{>}(E)\,\mathbf{G}^{<}(E)\,
  \mathbf{\Sigma}_{K_2}^{<}(E)\,\mathbf{G}^{>}(E)
 \right\},
 \nonumber
\end{align}
\end{widetext}
where $\mathbf{i}_K(E)$ is a matrix of the energy-resolved current operator
at interface $K$
\begin{equation}
 \mathbf{i}_K(E)\equiv\mathbf{\Sigma}_K^{<}(E)\,\mathbf{G}^{>}(E)
                     -\mathbf{\Sigma}_K^{>}(E)\,\mathbf{G}^{<}(E)
\end{equation}
The spin and charge noise at interface $K$ are expressed by
\begin{align}
 \label{SsKK}
 S^{(s)}_{KK} &= S_{K\uparrow,K\uparrow}-2S_{K\uparrow,K\downarrow}
               + S_{K_\downarrow,K\downarrow},
 \\
 \label{ScKK}
 S^{(c)}_{KK} &= S_{K\uparrow,K\uparrow}+2S_{K\uparrow,K\downarrow}
               + S_{K_\downarrow,K\downarrow},
\end{align}
where the spin resolved elements of the noise $S_{K\sigma,K\sigma'}$ 
can be obtained from (\ref{SK1K2}) by generalizing the counting field to include
spin.

Higher order cumulants for a multi-level molecule can be similarly derived from (\ref{dU}) in a, cumbersome but, straightforward manner.

%%%%%%%%%%%%%%%%%%%%%%%%%%%%%%%%%%%%%%%%%%%%%%%%%%%%%%%%%%%%%%%%%%%%%%%%%%%%%%%
\section{\label{heat}Heating}
Within the model (Fig.~\ref{scheme}), the charge current is compensated at each 
interface so that only spins are transferred through the junction. 
Nevertheless, since electrons cross the molecule-contact
interfaces also in the spin field-effect transistor, the question of heating of the molecular vibrations is
still important. 
To estimate the \emph{temperature} of the vibrations we extend an approximate
scheme used by one of us in a previous publication.\cite{raman}
The essence of the approach is quasi-particle assumption used to describe
vibrational degrees of freedom (phonons). In this case the only relevant
characteristic of the vibration (besides frequency) is its average population,
while actual nonequilibrium distribution may be disregarded (density of
states is a delta-function).

At steady-state influx of energy through the molecule-contact
interfaces, $J^{in}$, is compensated by outflux, $J^{out}$, 
so that energy of the molecule does not change $J=J^{in}-J^{out}=0$. 
Since energy is carried by both electrons and phonons, we get
\begin{equation}
 \label{zero}
 J_e+J_{ph}=0.
\end{equation}
Within the quasi-particle approximation the latter for a vibration $\omega_i$ 
coupled to a thermal bath is\cite{heat} 
\begin{equation}
 \label{Jph}
 J_{ph} = \frac{\Omega(\omega_i)}{\hbar} 
          \omega_i\left[N_{BE}(\omega_i)-N_i\right],
\end{equation}
where $\omega_i$ and $N_i$ ($i=1,2$) is a vibration and the corresponding average population of molecule $i$, respectively, whereas $\Omega$ is the spectral function of the bath, and $N_{BE}(\omega)=[e^{\beta\omega}-1]^{-1}$
is the Bose-Einstein distribution.

The electronic energy flux for a system coupled to a set of baths $B$ can be written as
\begin{align}
\label{Je}
 J_e =&
 	 \sum_B \int_{-\infty}^{+\infty}\frac{dE}{2\pi\hbar}\, E
       \mbox{Tr}[\Sigma_B^{<}(E)G^{>}(E)
        - \Sigma_B^{>}(E)G^{<}(E)],
\end{align}
where $\Sigma_B$ is the electronic self-energy due to the coupling to bath $B$.
In our model we want to estimate the \emph{temperature} of two vibrations, thus, 
following the approximate procedure introduced in Ref.~\onlinecite{opt},
we split the system into two parts at the molecule-molecule interface.
In this way we can speak about electronic energy flux in two parts 
of our system. Each part has two interfaces: one between the molecule
and corresponding contact and the other between the molecules. 
The electronic self-energy due to the coupling to the contact is given by
the standard expression (\ref{SK}). The self-energy on the molecule-molecule
interface is given by (\ref{SE}).

We obtain the electronic flux in the molecule $i$ part of the system through the following procedure. We begin with the Hamiltonian $\hat {\bar H}$ which is transformed to the rotating reference frames of the magnetic fields, Eq.(\ref{rf}), however, without performing the small polaron transformation.
We, then, treat the electron-vibration coupling in
the Born approximation\cite{peaksdips}. Within the non-crossing 
and quasi-particle approximations we accordingly obtain the electronic flux
\begin{equation}
 \label{Je_Born}
 J_e = \omega_i M_i^2\left([N_i+1]I^{(-)}_i-N_iI^{(+)}_i\right)
\end{equation}
where
\begin{equation}
 I^{(\pm)}_i \equiv \sum_{\sigma_1,\sigma_2}\int_{-\infty}^{+\infty}
 \frac{dE}{2\pi}\, G^{<}_{\sigma_1\sigma_2}(E)
                \, G^{>}_{\sigma_1\sigma_2}(E\pm \omega_i)
\end{equation}
Using (\ref{Jph}) and (\ref{Je_Born}) in (\ref{zero}) yields
\begin{equation}
 \label{Ni}
 N_i = \frac{\Omega(\omega_i)N_{BE}(\omega_i)-M_i^2I^{(-)}_i}
            {\Omega(\omega_i)+M_i^2(I^{(+)}_i-I^{(-)}_i)}
\end{equation}
The vibrational \emph{temperature} is obtained from (\ref{Ni}) under  assumption that the vibrations are populated according to the Bose-Einstein distribution. We note that a more physically motivated procedure to introduce 
the \emph{temperature} of the molecular vibrations at non-equilibrium can be used,\cite{heat}
however, for demonstration purposes the simpler procedure described above 
suffices.

Atomic cooling caused by sub-resonance optical excitation 
is a well-known effect.\cite{cool_review}
In molecular junction the analogous detuning may lead to
cooling of the molecular vibration. 
Here, we utilize the sub-resonance detuning of the electric and magnetic field 
frequencies from the energy difference of the inter-molecular and
contact-molecule electronic transitions
as a mechanism to pump energy out of the molecular vibration.
Naturally, cooling is most efficient when the detuning frequency
coincides with the frequency of the vibration.
Additional possibilities to cool the current carrying molecular junction
are discussed in Ref.~\onlinecite{cool}.

%%%%%%%%%%%%%%%%%%%%%%%%%%%%%%%%%%%%%%%%%%%%%%%%%%%%%%%%%%%%%%%%%%%%%%%%%%%%%%%
\section{\label{numres}Numerical results}
Here we present numerical results in order to illustrate the transport properties of 
our model for the spin field-effect transistor sketched in Fig.~\ref{scheme}.
We use the level width $\Gamma$ due to coupling to a contact as unit of energy.
Unless explicitly specified otherwise, the parameters for calculations are
temperature in the leads $T_e=0.3$, the positions of the levels in absence of
external fields $\varepsilon_1=\varepsilon_2=0$, escape rates to the contacts
(wide-band approximation is employed)
$\Gamma^{(K)}_{\sigma}=1$,
strength of coupling to external {\em dc} magnetic field
$\omega_0\equiv g\mu_B B^{(dc)}=-10$, strength of the coupling to the external
{\em ac} magnetic field $g\mu_B B^{(ac)}=0.01$ and its frequency
$\omega_B=20$, strength of the coupling to the external electric field
$V_E=0.5$. The parameters for the molecular vibrations are indicated in
each case separately.

\begin{figure}[t]%[htbp]
\centering\includegraphics[width=\linewidth]{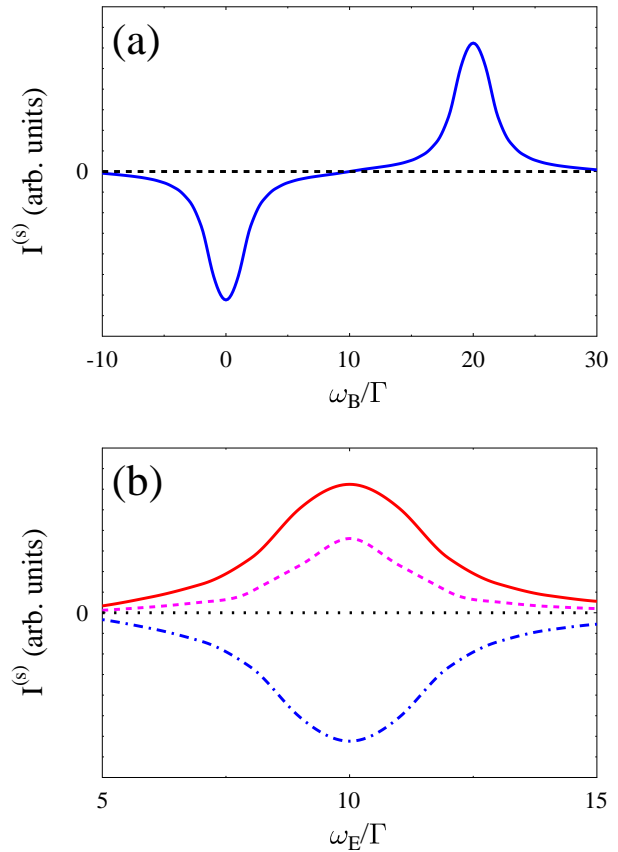}
\caption{\label{elast}
(Color online)
Elastic transport. Spin current vs. (a) {\em ac} magnetic field frequency
$\omega_B$ (solid line, blue) and (b) electric field frequency $\omega_E$.
The latter shows results for $V_E=0$ (dotted line, black),
$0.3$ (dashed line, magenta), and $0.5$ (solid line, red) at
$\omega_B=20$. Also shown curve for $V_E=0.5$ at
$\omega_B=0$ (dash-dotted line, blue). 
Dashed line (black) in (a) shows charge current.
See text for other parameters.
}
\end{figure}

\begin{figure}[t]%[htbp]
\centering\includegraphics[width=\linewidth]{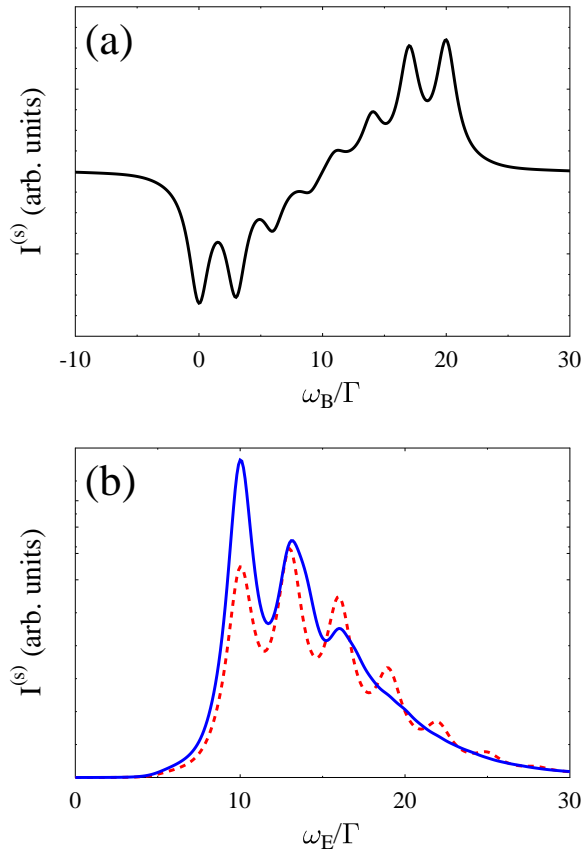}
\caption{\label{inelast}
(Color online)
Inelastic transport. Spin current vs. (a) {\em ac} magnetic field frequency
$\omega_B$ and (b) electric field frequency $\omega_E$.
The latter shows results for symmetric $\omega_1=\omega_2=3$ (dashed line, red)
and asymmetric $\omega_1=3$, $\omega_2=4$ (solid line, blue) cases.
Electron-vibration coupling is $M_1=M_2=3$.
Other parameters are as in Fig.~\ref{elast}.
}
\end{figure}

Figure~\ref{elast} shows the spin current $I^{(s)}$ in the elastic transport regime.
In the calculations, the shift of the levels due to changes in $\omega_B$ is assumed 
to be compensated by the {\em dc} magnetic field, so that levels 
are set as in Fig.~\ref{scheme} and do not move.
The spin current (solid line) vs. the external {\em ac} field frequency 
(spin bias) is plotted in Fig.~\ref{elast}a. 
Note that the electric field is an additional source of energy in the model, 
so that one will observe spin current even without spin bias in the contacts. 
The sign of the spin current shows the direction of the spin-up flux
(direction from left to right is chosen as positive). The flux switches
around $\omega_B/\Gamma=10$, as is shown in Fig.~\ref{elast}a.
The point of sign
change is defined by a competition between the spin bias and the electric field
mediated inter-molecular transfer. At $\omega_B=0$ the lower spin levels
are occupied, while the higher levels are empty (see Fig.~\ref{scheme}). 
In this regime
the electric field facilitates the spin-up flux from right to left, which
defines a negative sign of the spin flux. The spin bias grows with the frequency $\omega_B$, and the value $\omega_B/\Gamma=10$ the two processes cancel
each other. When the spin bias grows further, the population of the levels changes.
The higher levels become populated due to the increased bias, while the population of lower
levels diminishes (the level goes above the corresponding spin-resolved
chemical potential, see Fig.~\ref{scheme}). The positions of the peaks
are defined by the resonance condition for inter-molecule electron transfer: 
the position of the molecular levels is kept fixed by the {\em dc} field adjustment, 
while the frequency of the transition changes with both electric field
frequency $\omega_E$ and {\em ac} magnetic field frequency $\omega_B$,
see Eq.(\ref{Vbar}). The condition for the resonance is
$\omega_E\pm\omega_B=|\bar\varepsilon_{1\sigma}-\bar\varepsilon_{2\sigma}|$.
We will return to the question of the
role the spin bias plays in our discussion of the noise and heating properties. 
Note also, that the charge current (dashed line in Fig.~\ref{elast}a)
is identically zero (the flux of spin up electrons from left to right 
is compensated by the flux of spin down electrons from right to left).

Fig.~\ref{elast}b shows the dependence of the spin current on frequency of the external
electric field. Naturally, the dependence has a maximum at resonance. Shown are plots for magnetic field
frequency $\omega_B$ in the first (dash-dotted line) and second
(solid line) maximum of Fig.~\ref{elast}a.
Also shown are curves for smaller (dashed line) and zero coupling to
the electric field (dotted). In the absence of coupling no
spin current is observed. This indicates that, for the parameter range chosen, we are not in the regime of the spin pump described in 
Ref.~\onlinecite{Guo_spinpump}.

\begin{figure}[t]%[htbp]
\centering\includegraphics[width=\linewidth]{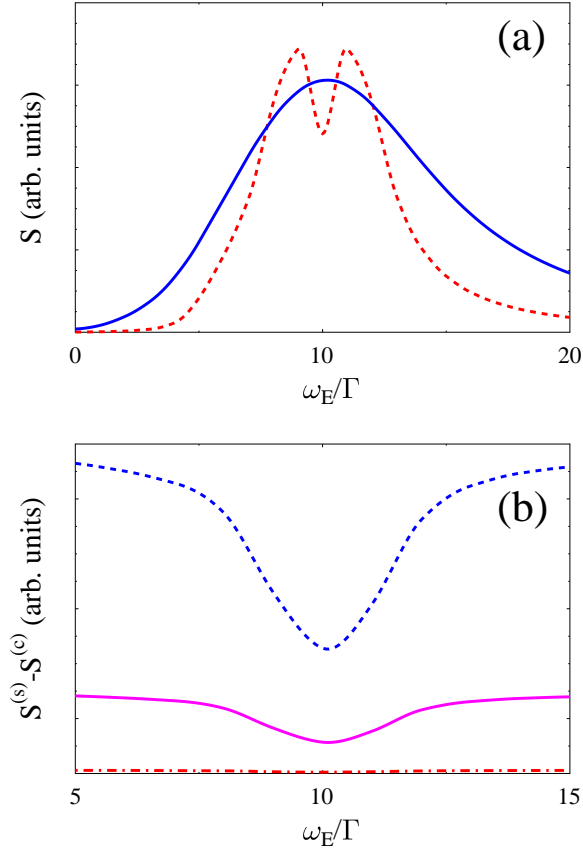}
\caption{\label{noise}
(Color online)
Noise properties.
(a) Zero-frequency charge and spin noise 
for  $\Gamma^{(K)}_\sigma=\Gamma$ (dashed line, red) and
$\Gamma^{(K)}_\sigma=5$ (solid line, blue).
vs. electric field frequency $\omega_E$.
(b) Difference between spin and charge zero-frequency noise for
$g\mu_B B^{(ac)}=0.01$ (dash-dotted line, red), $0.05$ (solid line, magenta),
and $0.1$ (dashed line, blue). See text for parameters.
}
\end{figure}

Figure~\ref{inelast} is inelastic analog of results presented in
Fig.~\ref{elast}.
Molecular vibrations are taken into account employing small polaron
transformation, and assuming separation of timescales, so that
introduction of the Franck-Condon factors becomes possible.
Fig.~\ref{inelast}a shows spin current vs. frequency
of {\em ac} magnetic field. Parameters for molecular vibrations are
$\omega_1=\omega_2=2$, electron-vibration coupling is $M_1=M_2=2$.
Other parameters are as in Fig.~\ref{elast}.
An unusual form of the vibration sidebands is caused by
resonance condition for intermolecular electron transfer
rather than resonance situation at molecule-contact interface.
This makes vibrational characteristic different from the one
presented in Ref.~\onlinecite{FranssonZhu_rf}.
At the same time, vibrational structure observed in spin current
vs. electric field frequency resembles such for charge current vs.
bias plots. Figure~\ref{inelast}b shows two such characteristics:
for symmetric $\omega_1=\omega_2=3$ (dashed line) and
asymmetric $\omega_1<\omega_2$ ($\omega_1=3$ and $\omega_2=4$, solid line) 
cases. Presence of a higher
frequency naturally leads to observation of less vibrational sidebands.

Figure~\ref{noise} displays the noise properties of the junction.
We show the results for elastic transport only, leaving the study of 
inelastic noise properties for a future publication. 
Fig.~\ref{noise}a shows the spin (or charge) zero-frequency noise 
as function of the external electric field frequency. Note that
while the charge current is identically zero, charge noise does exist.
Shown are the results for two situations: the spin levels of each molecule
are well resolved (dashed line, red) vs. essential overlap between the two
(solid line, blue). The first case corresponds to a situation with
well pronounced resonance behavior. In such situations, the transmission
probability at resonance approaches unity, which leads to suppression of 
the noise.\cite{Buttiker} The latter case corresponds to a situation
when the transmission probability is lower. In this case no noise suppression is observed at resonance. The effect is similar to the behavior of a
molecular junction in symmetric vs. asymmetric 
couplings at the two sides of the junction.\cite{noise}  

In contrast to charge noise, the spin noise strongly depends on spin-flip events 
within the system.\cite{Belzig_noise}
Fig.~\ref{noise}b shows the difference between the spin and charge zero-frequency 
noise at several magnitudes of the coupling to the external {\em ac} magnetic field.
Note that at resonance, where the electron transport through the junction
becomes pronounced, the difference between the spin and charge components diminishes.
 
\begin{figure}[t]%[htbp]
\centering\includegraphics[width=\linewidth]{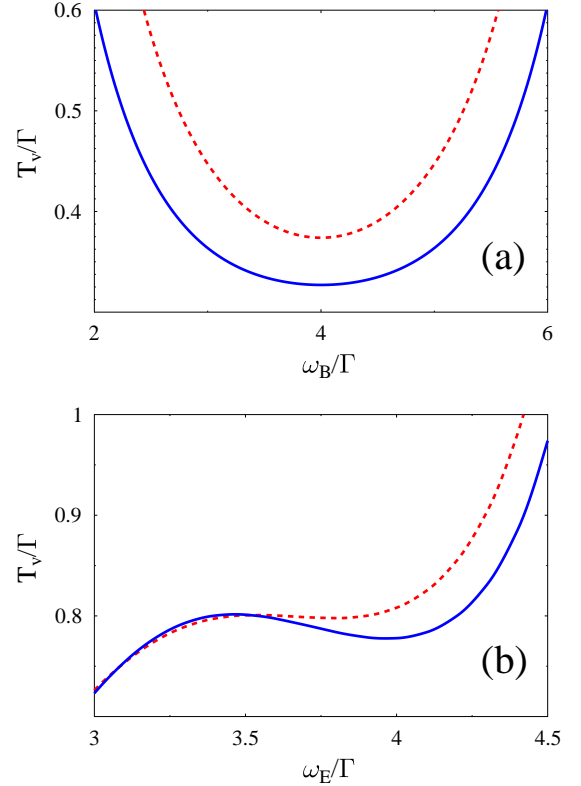}
\caption{\label{tv}
(Color online)
`Temperature' of the first $\omega_1$ (dashed line, red)
and second $\omega_2$ (solid line, blue) molecular vibration
vs. (a) {\em ac} magnetic field frequency $\omega_B$ 
and (b) external electric field frequency $\omega_E$.
See text for parameters.
}
\end{figure}

As was mentioned already earlier, there is no net charge transport between the leads. Nevertheless, 
charge transfer between contacts and molecule does occur.
Thus, the spin field-effect transistor may be heated.
As was discussed in the literature (see e.g. Refs. ~\onlinecite{Tao,cool}),
charge transfer by itself does not necessarily lead to heating
of the molecular device. Here, we demonstrate one such possibility.
Cooling of a molecular vibration is caused by tuning the external field 
frequency out of resonance (an analog of atomic optical cooling).
Figure~\ref{tv}a shows the \emph{temperatures} of the first $\omega_1$ (dashed line)
and second $\omega_2$ (solid line) molecular vibrations as functions
of the spin bias. The coupling to the {\em dc} magnetic field is $\omega_0=-4$, whereas the
frequency of the electric field is $\omega_E=4$. The molecular vibrations
are taken to be $\omega_1=1$ and $\omega_2=0.5$ with electron-vibration couplings
$M_1=M_2=0.5$. Other parameters of the calculation are introduced
at the start of the section. 
The parameters (electric and magnetic field frequencies)
are chosen in such way that the most probable
(resonant) electronic tunneling through the junction has to go uphill
in energy at the contact-molecule and inter-molecule transition steps.
In this case, the tunneling energy difference required for resonant
electron transition is taken from the molecular vibration, which leads
to cooling of the device.
Fig.~\ref{tv}b shows similar behavior when the frequency of the electric field
is used as a control instead. The molecular vibrations are 
$\omega_1=0.5$ and $\omega_2=0.2$. Here cooling takes place at the 
molecule-molecule interface only and is not as effective.
However, one can achieve a stronger signal in this case. 
The efficiency of the cooling at maximum spin current is an interesting 
question for future studies.

%%%%%%%%%%%%%%%%%%%%%%%%%%%%%%%%%%%%%%%%%%%%%%%%%%%%%%%%%%%%%%%%%%%%%%%
\section{\label{conclude}Conclusion}
Spintronics is a quickly developing field of research (both experimental and
theoretical). Here, we studied the spin current transport properties within
a model for an all-electric controlled spin field-effect transistor. 
The model provides pure spin currents through a junction which consists of 
two spin-pumps (molecules under influence of external {\em dc} magnetic 
fields)\cite{Guo_spinpump} each attached to its normal metal contact. 
The two parts are coupled by an external electric field, which serves as a 
control for the spin flux. Within a full counting statistics approach
to multi-level systems, we discuss elastic and inelastic spin flux
and noise properties of the junction. External electric and magnetic
fields are indicated as possible controls of the spin current through
the junction. The charge and spin zero-frequency noises
are shown to be different when spin-flip processes within the junction 
become pronounced. Zero-frequency spin noise as function of the external 
electric field shows the same single to double peak structure transition
when the tunneling probability approaches unity. Similar behavior was observed 
earlier for a model of charge field-effect transistor.\cite{noise} 

Problem of molecular junction heating recently discussed in the literature in
connection to charge transport, retains its importance also for
spin molecular devices. Within the model, we discuss possibility to use external
fields slightly de-tuned from the molecular resonances for cooling the
molecular vibrations. This process is similar to optical cooling of atoms.
We find that external {\em ac} magnetic fields may be effective cooling media.
External electric fields also provides the effect. The efficiency of the 
cooling at maximum spin flux is an interesting question for future studies.

%%%%%%%%%%%%%%%%%%%%%%%%%%%%%%%%%%%%%%%%%%%%%%%%%%%%%%%%%%%%%%%%%%%%%%%%
%
\begin{acknowledgments}
J.F. thanks UCSD for its hospitality during his visit in July 2009, and the Swedish Research Council and the Royal Swedish Academy of Sciences for financial support. 
M.G. gratefully acknowledges support by the UCSD (startup funds), 
the UC Academic Senate (research grant), 
and the U.S.-Israel Binational Science Foundation. 
\end{acknowledgments}

%%%%%%%%%%%%%%%%%%%%%%%%%%%%%%%%%%%%%%%%%%%%%%%%%%%%%%%%%%%%%%%%%%%%%%%%%
\appendix
\section{\label{appA}Derivation of Eq.(\ref{SE})}
Here we derive an expression (\ref{SE}) for the self-energy due to the coupling
to the external electric field within the rotating-wave approximation.
We start by partitioning the total Hamiltonian of the system into a zero-order 
Hamiltonian $\hat{\bar H}_0$ and perturbation $\hat{\bar V}$, Eq.(\ref{Vbar}). 
The first is given by an analog of Eq.(\ref{H0})
after transformation to rotating frames. It will 
be similar to (\ref{H0}) with molecular and contact states energies 
renormalized according to (\ref{eibar}) and (\ref{ekbar}), respectively, 
and without the time dependence of magnetic field terms.
For the moment, we disregard the coupling to molecular vibrations.
Hamiltonian $\hat{\bar H}_0$ defines the zero-order GFs
of the system $\mathbf{G}^{(0)}$. We treat the interaction $\hat{\bar V}$
by perturbation theory, i.e expanding the evolution operator on the Keldysh contour
(interaction representation) $\exp[-i\int_c d\tau \hat V_I(\tau)]$ to
second order in $V_E$. This leads to a Dyson type equation for the Green 
function
\begin{align}
 \label{G1}
 &\mathbf{G}(\tau,\tau') = \mathbf{G}^{(0)}(\tau,\tau')
 \\
 &+ \int_c d\tau_1\int_c d\tau_2\, \mathbf{G}^{(0)}(\tau,\tau_1)\,
   \mathbf{\Sigma}^{(E)}(\tau_1,\tau_2)\, \mathbf{G}^{(0)}(\tau_2,\tau')
 \nonumber
\end{align}
with matrix elements for the self-energy given by
\begin{align}
 \label{SE1}
 &\Sigma^{(E)}_{m\sigma_1,n\sigma_2}(\tau_1,\tau_2) =\delta_{m,n}
 (V_E e^{-i\omega_E t_1}+V_E^{*}e^{i\omega_E t_1})
 \nonumber \\
 &\times (V_E e^{-i\omega_E t_2}+V_E^{*}e^{i\omega_E t_2})
 \\
 &\times G^{(0)}_{\bar m\sigma_1,\bar m\sigma_2}(\tau_1,\tau_2)
 \exp[i(-1)^{m}\omega_B(\sigma_1 t_1-\sigma_2 t_2)]
 \nonumber
\end{align}
where $m,n=1,2$ numerate molecule in the junction, whereas $\bar m$ means
opposite of $m$.
Application of the rotating wave approximation leaves
only terms proportional to $|V_E|^2$ and enforces $\sigma_1=\sigma_2$, in (\ref{SE1}).
The resulting expression is presented in Eq.(\ref{SE}).

In the case of the small polaron transformation, the
coupling to molecular vibrations is treated by dressing the expression
(\ref{SE}) by Franck-Condon factors 
\begin{align}
 \label{SE_FC}
 &\Sigma_{m\sigma,n\sigma'}^{(E)}(\tau,\tau') = 
 \delta_{m,n}\delta_{\sigma,\sigma'}
 2 |V_E|^2 \cos\omega_E(t-t') 
 \nonumber \\ &\times
 G_{\bar m\sigma,\bar m\sigma}(\tau,\tau') e^{i(-1)^m\sigma\omega_B(t-t')} 
 \\ &\times
 \langle T_c \hat X_{\bar m}(\tau)\hat X_{\bar m}^\dagger(\tau')\rangle
 \langle T_c \hat X_m(\tau')\hat X_m^\dagger(\tau)\rangle
 \nonumber
\end{align}
When perturbation theory is employed, the GF in Eq. (\ref{G1}) 
is considered as a zero-order GF of the Born approximation.

%%%%%%%%%%%%%%%%%%%%%%%%%%%%%%%%%%%%%%%%%%%%%%%%%%%%%%%%%%%%%%%%%%%%%%%%%%%%%%

%%%%%%%%%%%%%%%%%%%%%%%%%%%%%%%%%%%%%%%%%%%%%%%%%%%%%%%%%%%%%%%%%%%%%%%%%%%%%%

\begin{thebibliography}{99}
\bibitem{footnote_2}Deriving (\ref{SK1K2}) from (\ref{dU})
   we use mathematical matrix identity 
   $\frac{d\mathbf{M}^{-1}(x)}{dx}=-\mathbf{M}^{-1}(x)
    \frac{d\mathbf{M}(x)}{dx}\mathbf{M}^{-1}(x)$
\end{thebibliography}
\end{document}